\documentclass[onecolumn]{aastex631}

\usepackage{amsmath}	
\usepackage{amssymb}	
\usepackage{enumitem}
\usepackage{verbatim} 
\usepackage[caption=false]{subfig}
\usepackage{array}

\usepackage{listings}
\usepackage{hyperref}

\begin{document}

\makeatletter\let\frontmatter@title@above=\relax

\newcommand{\todo}[1]{\textcolor{red}{ToDo: #1}}

\email{tdaylan@princeton.edu}
\author[0000-0002-6939-9211]{Tansu Daylan}
\affiliation{Department of Astrophysical Sciences, Princeton University, 4 Ivy Lane, Princeton, NJ 08544, USA}
\affiliation{LSSTC Catalyst Fellow}

\author[0000-0003-3195-5507]{Simon Birrer}
\affiliation{Department of Physics and Astronomy, Stony Brook University, Stony Brook, NY 11794, USA}

\begin{abstract}

We recommend a deeper extension to the High-Latitute Wide Area Survey planned to be conducted by the Nancy Grace Roman Space Telescope (\emph{Roman}). While this deeper-tier survey extension can support a range of astrophysical investigations, it is particularly well suited to characterize the dark matter substructure in galactic halos and reveal the microphysics of dark matter through gravitational lensing. We quantify the expected yield of \emph{Roman} for finding galaxy-galaxy-type gravitational lenses and motivate observational choices to optimize the \emph{Roman} core community surveys for studying dark matter substructure.  In the proposed survey, we expect to find, on average, one strong lens with a characterizable substructure per \emph{Roman} tile (0.28 squared degrees), yielding approximately 500 such high-quality lenses.
With such a deeper legacy survey, \emph{Roman} will outperform any current and planned telescope within the next decade in its potential to characterize the concentration and abundance of dark matter subhalos in the mass range 10$^7$-10$^{11}$\,M$_{\odot}$.
\end{abstract}

\begin{center}
\textbf{Searching for dark matter substructure: a deeper wide-area community survey for \emph{Roman}} \\
\vspace{1em}
\noindent\textbf{Roman Core Community Survey (CCS) White Paper} \\
\textbf{CCS:} High-Latitude Wide Area Survey \\
\textbf{Scientific Categories:} galaxies; large-scale structure of the universe\\
\textbf{Additional scientific keywords:} Cosmology, dark matter distribution, gravitational lensing\\
\textbf{Submitting Author:} Tansu Daylan (Princeton University), tdaylan@princeton.edu

\end{center}


\section{Investigating Dark Matter Substructure Using Gravitational Imaging}
\label{sect:Introduction}

The hierarchical, bottom-up nature of structure formation in the Universe suggests that dark matter halos must contain a fraction of their mass in the form of previously accreted subhalos that survive evaporation and tidal dissipation. In the Cold Dark Matter \citep[CDM;][]{Navarro1996, Bullock2001} model, these subhalos constitute $\sim$1\% of the halo mass and can individually have masses as low as Earth mass. Departures from the CDM paradigm typically predict a cutoff in the subhalo mass function. Therefore, the dark matter subhalo mass function is a promising observable capable of distinguishing between dark matter models \citep{Bechtol2022}.

\begin{figure*}
    \centering
    \includegraphics[width=\textwidth]{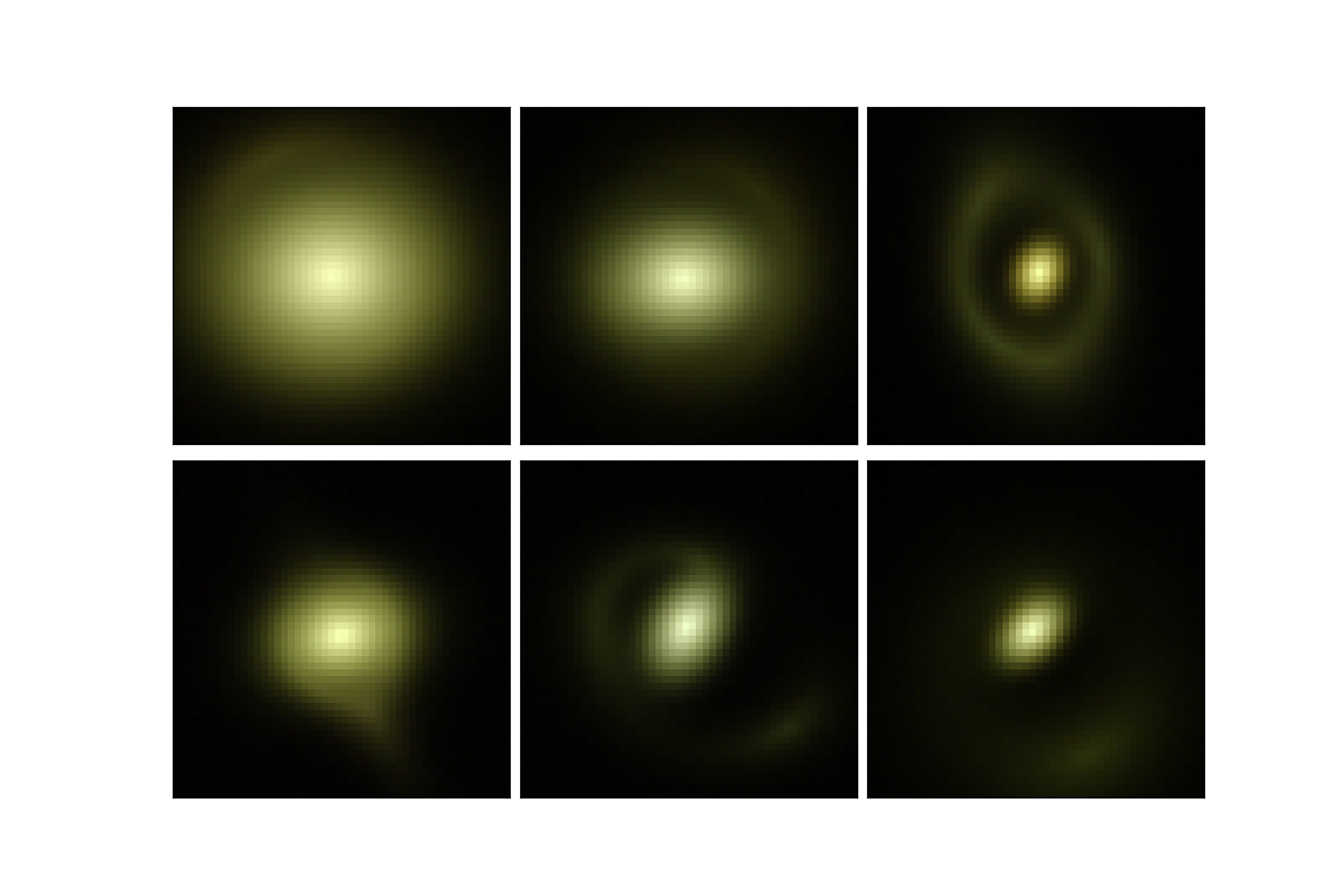}
    \caption{A selected subset of simulated \emph{Roman} images of galaxy-galaxy-type strong lenses, which are amenable to substructure characterization. The brightness of the images has been stretched using the \texttt{arcsinh} function to amplify the faint features such as lensed arcs and arclets. The \emph{Roman} bandpasses F184, F129, and F062 have been mapped to the red, green, and blue channels. One side of each image corresponds to five arcseconds.}
    \label{figure: grid}
\end{figure*}

Subhalos with masses less than 10$^7$\,M$_{\odot}$ do not form stars and, being dark, are only observationally accessible via gravitational interactions. Strong lensing of a background quasar or galaxy by another galaxy in the foreground, i.e., a galaxy-galaxy-type strong lens, is a promising tool to study the mass, size, and concentration of dark matter halos and reveal their substructure. In particular, strong lensing of a late-type background galaxy into multiple, magnified arcs and arclets can enable the measurement of the substructure in the foreground early-type galaxy and small-scale halo abundances along the line of sight. Several simulated strong lenses are shown in Figure~\ref{figure: grid}.

The imaging data collected by the Hubble Space Telescope (HST) on galaxy-galaxy-type lenses have been restricted to $\sim$100 grade A lenses. Because of the small field of view of HST, detections were made possible by defining candidates based on a compound spectroscopic signature in the Sloan Digital Sky Survey \citep{Bolton2005}. Several surveys conducted with HST are the Sloan Lens ACS Survey \citep[SLACS;][]{Bolton2008, Auger2009}, the Sloan WFC Edge-on Late-type Lens Survey \citep[SWELLS;][]{Treu2011, Brewer2012} and the BOSS Emission-Line Lens Survey \citep[BELLS;][]{Brownstein2012}. SLACS was a particularly deep survey, achieving $\sim$400 seconds of exposure time per band and a total exposure time of $\sim$one HST orbit ($\sim$90 minutes) for some strong lenses.

Building on this legacy, the Nancy Grace Roman Space Telescope \citep[\emph{Roman};][]{Spergel2015} will revolutionize our ability to detect and characterize strong lenses using various approaches to modeling these strong lens systems, including forward modeling \citep{Vegetti2010, Birrer2021, Daylan2018} and neural networks \citep{Wagner-Carena2023}. \emph{Roman} will be able to significantly improve the tentative constraints on the dark matter subhalo mass function achieved by the modeling of the available lenses \citep{Vegetti2014}. The key features of \emph{Roman} for studying dark matter substructure will be its large field of view of 0.28 squared degrees, effective area of $\sim$2.5 squared meters, and high angular resolution with a point-spread function (PSF) full-width at half maximum of $\sim$0.1 arcsecond. Furthermore, its available passbands will cover 0.5 to 2.3 microns, and a limiting AB magnitude of $\sim$25 will be achieved across the passbands with a single exposure of 55 seconds.

\textit{Roman} is expected to conduct three core community surveys: the High-Latitude Wide Area Survey (HLWAS), High-Latitude Time Domain Survey (HLTDS), and the Galactic Bulge Time Domain Survey. In addition, a shallower full-sky survey is also being proposed as part of this core community survey. In particular, the HLWAS is expected to target a wide ($\sim$1700 squared degrees) area at high galactic latitudes using $\sim$150 second exposures, several filters, and multiple exposures for each filter to achieve dithering. While the two-year HLWAS is expected to largely be an imaging survey, it will also have a smaller slitless spectroscopy component using the grism. In contrast, the HLTDS is expected to cover a small, $\sim$5 squared degree field and go much deeper using six months of \emph{Roman} time.

The characterizability of a strong lens is proportional to the brightness of the magnified images of the source galaxy and their separation from the surface brightness of the foreground galaxy. Most importantly, the extreme rarity of these favorable conditions makes the survey area a critical factor for finding strong lenses. Because of their isotropic distribution, the number of galaxy-galaxy-type strong lenses is expected to scale linearly with the survey area, which motivates wide-area surveys to find them. This renders HLTDS an unfavorable option for finding a large number of new strong lenses. In contrast, given its much wider area, HLWAS will produce a great data set for finding galaxy-galaxy-type strong lenses. However, the characterization of dark matter substructure requires high signal-to-noise imaging data such that the lowest mass of the detectable subhalo decreases with the signal-to-noise (SNR) \citep{Despali2022}. \textbf{Consequently, HLWAS will not be sensitive to subhalos lower than $\sim$10$^{10}$M$_{\odot}$, leaving out a fast discovery potential of \emph{Roman}.}

\section{The Deep High-Latitude Survey}
\label{section: HLDS}

\textbf{We recommend the inclusion of a smaller and deeper tier within the HLWAS, the High-Latitude Deep Survey (HLDS), with an exposure time of $\sim$21,600 seconds (6 hours) per tile over an area of $\sim$100 squared degrees at high galactic latitudes}. Furthermore, we recommend the choice of two short-wavelength passbands for these observations: F087 and F106. Even though shorter wavelength observations generally provide the most opportunistic angular resolution and sensitivity for the bluer late-type galaxies in the background, the \emph{Roman} PSF will be undersampled blueward of $\sim$1.2 micron, which will make it particularly challenging to model and interpret F062 observations at small angular scales. The sky position and shape of this deeper HLWAS tier are flexible to characterize strong lenses at the level needed to measure dark matter substructure as long as the tier is situated away from the galactic and equatorial planes. Nevertheless, choosing the footprint to align with deep imaging surveys, such as the Hyper Suprime-Cam Subaru Strategic Program \citep[HSC-SSP;][]{Aihara2018} wide field with a limiting r magnitude of 26, can improve prospects for the lensed source characterization. The total exposure time of 21,600 seconds would be broken down into two bands and three dithered pointings per band, similar to HLWAS. Because strong lenses are static objects, the number and the temporal footprint of the observations would be free to be optimized for other science cases. A regular cadence would be desirable to maximize the prospects of discovering lensed supernovae or quasars. Alternatively, multiple exposures could be taken consecutively to reduce time lost to slew and settle, yielding a highly efficient survey.

\begin{figure}
    \centering
    \includegraphics[width=1\textwidth]{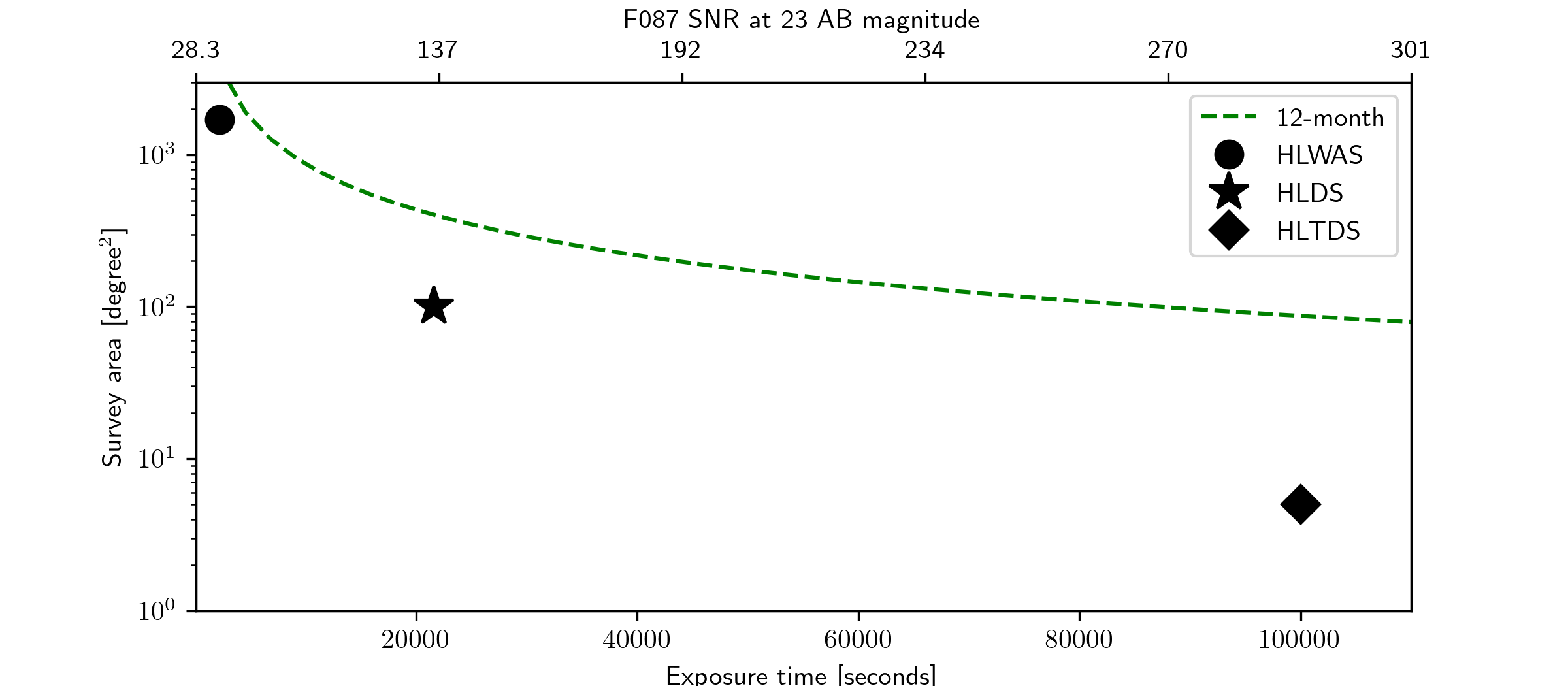}
    \caption{The tradeoff between the exposure time and the survey area. To guide the eye, the dashed green line indicates a hypothetical 12-month survey with a single exposure with one filter over a tile with no time lost to the readout, slew, or settle. The nominal HLWAS, HLTDS, and our proposed HLDS are shown with a black circle, diamond, and star, respectively. The HLDS falls between the HLWAS and HLTDS regarding its depth and survey area, filling a significant observational gap for \emph{Roman}.}
    \label{figure: AreaExposure}
\end{figure}

As shown in Figure~\ref{figure: AreaExposure}, while deep surveys provide the highest SNR measurements and observational access to the faintest and farthest objects, wide surveys yield a statistically more robust larger sample of objects and opportunities to study their spatial distribution. However, a healthy use of \emph{Roman} time will require a broad distribution over survey depth and area via the use of intermediate-depth surveys. On the other hand, it is plausible that a large fraction of the Guest Observer (GO) proposals will have a particular target list focused on a specific science goal. Our recommendation for a middle-tier survey aims to fill this gap between wide-area surveys and the deep drilling fields that may be proposed as part of the GO program. In addition, the HLDS would be a legacy survey for the entire strong lensing field, being able to characterize precisely-measured and representative samples of strong lenses that can be used for cosmographic studies, galaxy evolution, and statistical explorations that are currently yet unexplored \citep[e.g.][]{Birrer2018}.

\subsection{Simulating \textit{Roman}'s view of the galaxy-galaxy-type lens population}

\begin{figure}
    \includegraphics[width=0.48\textwidth]{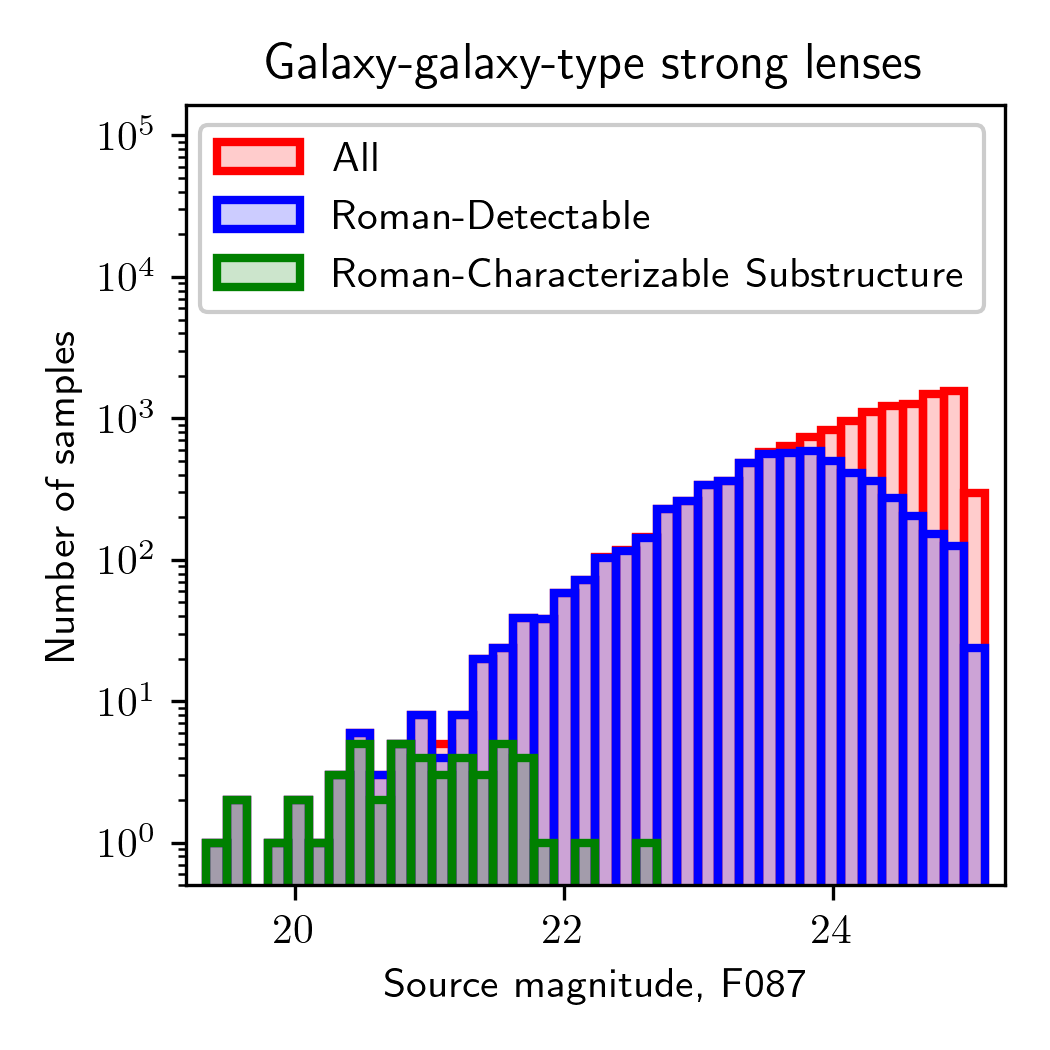}
    \includegraphics[width=0.48\textwidth]{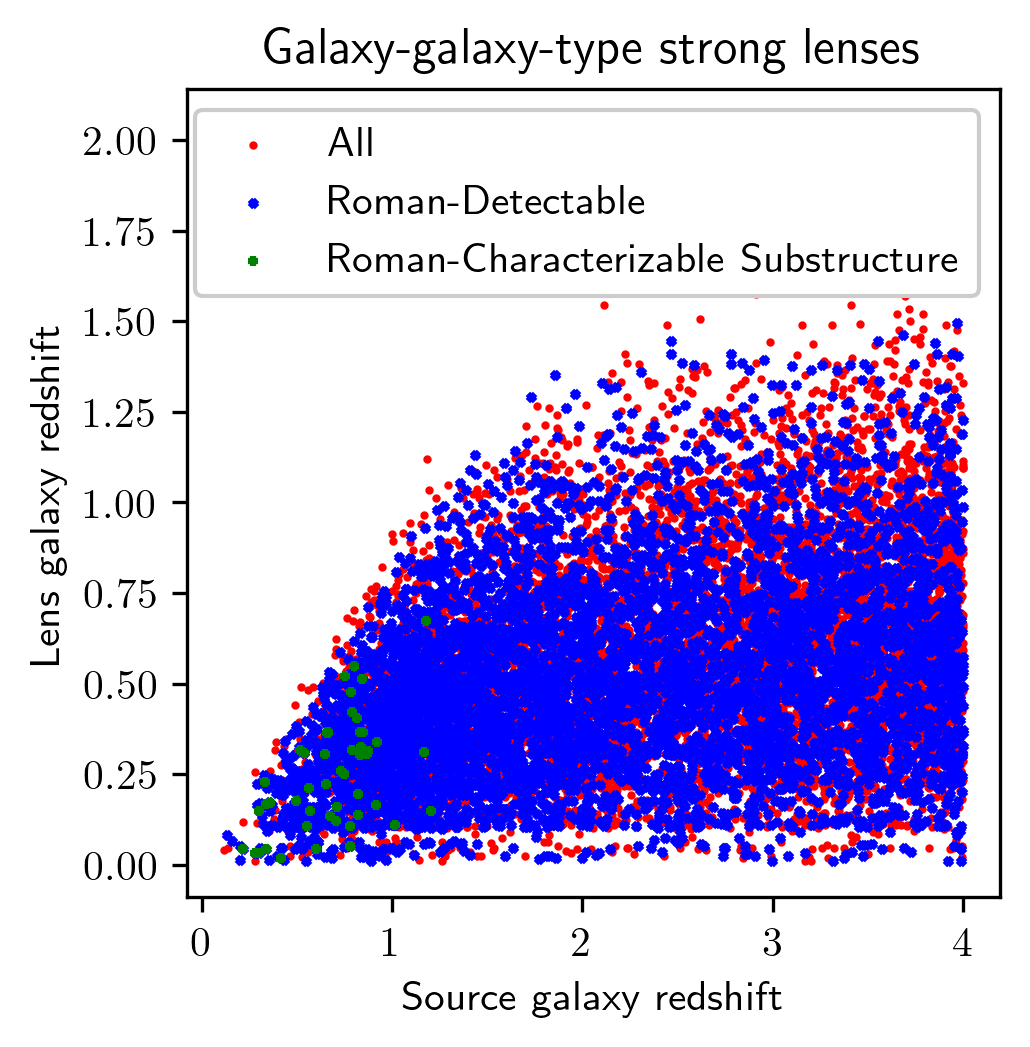}
    \caption{Properties of the galaxy-galaxy-type strong lenses in a simulation over two squared degrees. Left: histograms of all simulated galaxy-galaxy-type strong lenses (red), those that are detectable by \emph{Roman} (blue), and those that are amenable to the characterization of substructure via the detection of low-mass subhalos (green). Right: a scatter of lens and source galaxy redshifts for the same classes of simulated galaxy-galaxy-type lenses. The green points overlap the underlying blue points, which in turn overlap the underlying red points.}
    \label{figure: yield}
\end{figure}

To estimate \emph{Roman}'s yield for galaxy-galaxy-type strong lenses, we simulated a population of galaxy-galaxy-type lenses by randomly drawing galaxies early- and late-type galaxies across cosmic time using \textsc{skypy}'s \citep{Amara2021} galaxy population model and then constructed strong lens systems by requiring positional alignment between the lens and source galaxy populations with an Einstein radius, sufficient to produce at least two images. We then checked what fraction of these lenses would be observationally detectable by Roman by taking magnified source F087 magnitudes lower than 21 and a maximum distance between the images greater than 0.8 arcseconds. This produces approximately 80,000 high-significance galaxy-galaxy-type strong lens detections within the nominal 1700 squared degrees of HLWAS. We further find that the HLDS would yield approximately 500 strong lenses for which the detection of low-mass subhalos will be accessible. The simulation results are shown in Figure~\ref{figure: yield}.

\section{Conclusion}
\label{section: Conclusion}

In this work, we recommended a deeper tier of observations within \emph{Roman's} HLWAS optimized to achieve precision cosmology using strong lenses. Our recommendations for the core community survey definition process are motivated to optimize \emph{Roman}'s discovery potential and scientific return focused on dark matter investigations. Our proposed HLDS is designed to generate a wide and deep set of imaging data unique in its resulting combination of SNR and angular resolution, which can yield a large number of characterizable strong lenses with unique discovery potential compared to current and near-future observatories.

\section{Acknowledgements}

This work was supported by an LSSTC Catalyst Fellowship awarded by LSST Corporation to T.D. with funding from the John Templeton Foundation grant I.D. \#62192.

\software{\textsc{SkyPy}\citep{Amara2021}, \textsc{lenstronomy}\citep{Birrer2021},  \textsc{sim-pipeline}, LSST DESC-SLSC simulation pipeline (DESC and SLSC Collaboration, in prep}).

\bibliography{references}

\end{document}